\documentclass[prb,reprint,aps]{revtex4}
\usepackage{graphicx}

\begin{document}
\title
{Electronic Structure of Te and As Covered Si(211)}

\author{Prasenjit Sen, Inder P. Batra, S. Sivananthan, and C.H. Grein}
\affiliation{
Department of Physics,
University of Illinois at Chicago, Chicago, Illinois 60607-7059}

\author{Nibir Dhar}
\affiliation{
US Army Research Laboratory,
Adelphi, Maryland 20783-1197}

\author{S. Ciraci}
\affiliation{
Department of Physics, Bilkent University,
Ankara 06533, Turkey}

\date{\today}

\begin{abstract}
Electronic and atomic structures of the clean, and As and Te covered
Si(211) surface are studied using pseudopotential density functional
method.
The clean surface is found to have $(2\times 1)$ and rebonded
$(1\times 1)$ reconstructions as stable surface structures, 
but no
$\pi$-bonded chain reconstruction. Binding energies of As and Te
adatoms at a number of symmetry sites on the
ideal and $(2\times 1)$ reconstructed surfaces have been calculated
because of their importance in the epitaxial growth of CdTe and other
materials on the Si(211) surface. The special symmetry sites on these
surfaces having the highest binding energies for isolated As and Te
adatoms are identified. But more significantly, several sites are
found
to be nearly degenerate in binding energy values. This has important
consequences for epitaxial growth processes. Optimal structures 
calculated for 0.5 ML of As and Te coverage reveal that the
As adatoms dimerize on the surface while the Te adatoms do not.
However,
both As and Te covered surfaces are found to be metallic in nature.

\end{abstract}


\maketitle

\section{Introduction}

Silicon surfaces have been the subject of intense theoretical and experimental
investigations. However, the high index surfaces have not received nearly as
much attention as the low index ones. 
The (001) and (111) surfaces of Si have been studied most extensively. 
Only recently, the (211), (311), (331) and other higher index surfaces 
have attracted some attention~\cite{chadi, olshanetsky}. The
emerging interest in the high Miller-index surfaces is primarily due to
the fact that these surfaces may play important role in technological applications.

\begin{figure}[h]
\includegraphics{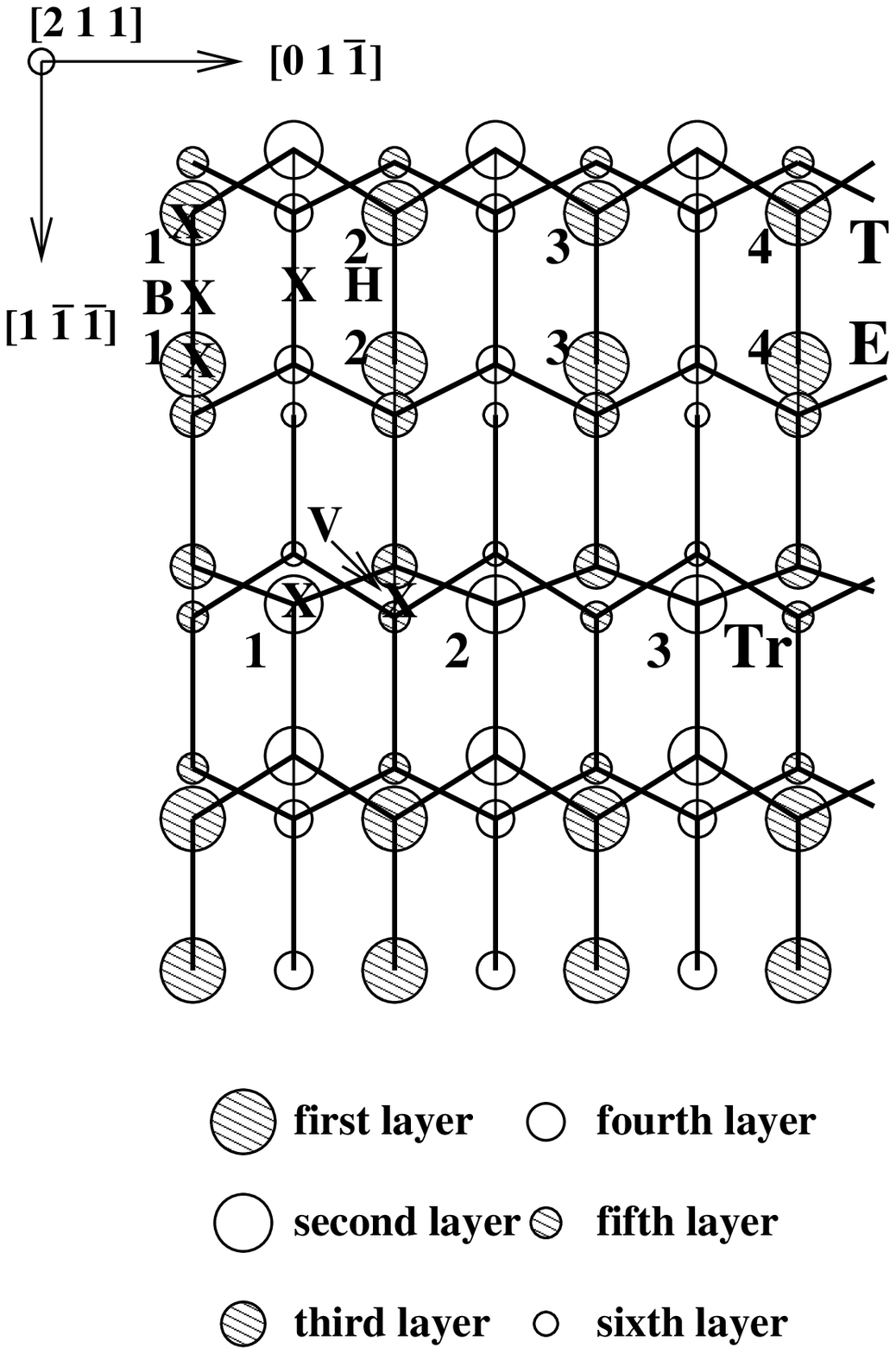}
\caption{Top view of an ideal bulk terminated Si(211) surface. The
top layer terrace and edge atoms and the second layer 
trench atoms are marked by {\bf T}, {\bf E}, and
{\bf Tr} respectively. The other symmetry sites at which adsoprtion of
As and Te is studied have also been marked (by {\bf X}).}
\label{fig:ideal}
\end{figure}

An interesting property of many of the higher index surfaces of Si is
the occurrence of steps and terraces. The Si(211) surface, for example,
can be looked at as stepped arrangement of narrow (111) terraces. 
A top view of the ideal Si(211) surface is shown in
Fig.~\ref{fig:ideal}. The surface consists of two-atom wide terraces
along [$1\bar{1} \bar{1}$]. Two consecutive terraces are separated by
steps and are 9.41 \AA~ apart in the [$1\bar{1} \bar{1}$] direction,
while they extend
infinitely along [$01\bar{1}$]. The atoms marked {\bf T} on the
terrace are three-fold coordinated and so has one dangling bond each,
while those on the step edge, marked {\bf E} are two-fold coordinated
and have two dangling bonds each. The behavior and possible
reconstructions of such stepped surfaces is of fundamental interest as
they can involve physics not seen in the lower index surfaces. On the
other hand, these high index surfaces can be the natural choice for
epitaxial growth of polar (both III-V and II-VI) semiconductors on a Si
substrate. As discussed~\cite{WKI} earlier, Si(211)
surface leads to better quality epitaxial growth of GaP as compared to
Si(001) because it satisfies both the requirements of interface
neutrality and offering inequivalent binding sites to Ga and P.  The Si(211) 
surface has atoms with both one and two dangling bonds.
The atoms with two dangling bonds can accomodate P, whereas Ga binds with Si(211)
that has a single dangling bond. Large area high quality CdTe layers have
also been grown on the Si(211) surface for subsequent growth of
HgCdTe.~\cite{rujirawat, yang}

One basic point to understand about a surface is its
possible reconstructions. There have been a few 
studies dealing with the reconstruction of Si(211). However, the
results have not been conclusive. Scanning tunneling microscopy
(STM) studies by Berghaus {\it et al.}~\cite{berghaus} reveal both
regions of $(2\times 1)$ reconstruction and missing edge atoms
possibly leading to a rebonded $(1\times 1)$ surface (discussed below).
Low-energy electron diffraction (LEED) and STM studies by Wang and
Weinberg~\cite{wang} reveal a weak $2\times $ reconstruction along the
$[0 1 \bar{1}]$ direction and a definite $2\times $ reconstruction along
$[1\bar{1} \bar{1}]$. ($2\times $ reconstruction here means a doubling
of the unit cell in the corresponding direction).
Olshanetsky and Mashanov observed~\cite{olshanetsky}
a $(4\times 2)$ structure in their LEED studies on the Si(211)
surface. Kaplan~\cite{kaplan} observed different
structures under different conditions of surface preparation. Wright
{\it et al.}
observed both $(1\times 2)$ and $(4\times 2)$ reconstructions of the
clean Si(211) surface annealed in vacuum.~\cite{WKI}

The properties of the clean surface being not fully understood as they
are,
there have been few studies of adsorption of other elements on the
Si(211) surface. Such studies are important in the context of
epitaxial growth of other materials on Si. Wright {\it et
al.}~\cite{WKI} found
both $(6\times 1)$ and $(6\times 2)$ reconstructions on Ga deposited
Si(211) surface, whereas Kaplan~\cite{kaplan} found only a $(6\times
1)$ pattern in his LEED studies of the same surface. Yang and
Williams~\cite{yangwil}, in their study on the effects of carbon 
contamination on the
Si(211) surface, found that while a small amount of C removes the
$(1\times 2)$ reconstruction of the clean surface and leads to a
$(4\times 2)$ reconstruction, larger quantities of C leads to
facetting. Michel {\it et al.} have studied Br adsorption on the
Si(211) surface~\cite{michel} by an X-ray standing wave technique.
They conclude that Br adsorption reverts the $(2\times 1)$
reconstruction of the clean surface into $(1\times 1)$ and that Br is
adsorbed at least at two different sites on the surface. A similar
conclusion is reached by Dhar {\it et al.}~\cite{dhar} in their study of Te
adsorption on the Si(211) surface. Te is found to adsorb at more than
one sites. This work also estimates the binding energy of Te on
Si(211) to be 3.46 eV/atom.

There have been even fewer theoretical studies on the Si(211) surface.
Chadi~\cite{chadi} studied the atomic structure of the (211), (311)
and (331) surfaces of Si by the tight-binding approach. The two first
principles calculations are by Grein~\cite{grein} and 
Mankefors~\cite{mankefors}.
Of these, the second one studies a system with only two atoms in each
(211) atomic plane (30 atomic planes containing 60 atoms). In view of
the fact that each terrace contains two inequivalent atoms,
this system cannot give any information about possible
reconstructions, as for example, a $2\times $ reconstruction along 
[$01\bar{1}$] would require at least two (equivalent) edge atoms.
To our knowledge, there has also been no theoretical study of adsorption of
any material on the Si(211) surface.
For epitaxial growth of various materials on the
Si(001) surface ({\it e.g.}, Ge, which normally 
grows in the Stranski-Krastanov mode)
it has been found that As or Te act as surfactants, leading to nice 
layer-by-layer growth. It is expected that these materials can act as
good surfactants even on the Si(211) surface. Hence one would like to
understand the energetics and structure of As and Te adsorption on
Si(211). Moreover, structure of a Te layer on Si(211) is also
important in the context of CdTe growth on this surface. With these
applications in mind, and also to understand its structure and
electronic properties, we performed extensive first-principles 
calculations on the clean, and As and Te covered Si(211) surface.

We organize the rest of the paper in the following way.
Section~\ref{sec:method} gives the parameters we used in the 
pseudopotential density functional calculations.
Section~\ref{sec:results} discusses the results of our calculation.
Subsection~\ref{subsec:clean} gives the results for the clean surface
while~\ref{subsec:As} and~\ref{subsec:Te} give results of As and Te
adsorption on the ideal and $(2\times 1)$ reconstructed surfaces. Finally,
in the closing section~\ref{sec:conclusion}, we state our main conclusions.

\section{Method}
\label{sec:method}
First principles calculations are carried out within the density
functional theory (DFT). The Si(211) surface is represented in a
repeated slab geometry. Each slab contains seven Si(211) layers and a
12 \AA~ vacuum region. Each layer contains 8 Si atoms--2 along 
[$1\bar{1} \bar{1}$] and 4 along [$01\bar{1}$]. In other words, each
layer contains 4 edge and 4 terrace Si atoms. The Si atoms in the bottom
layer have their dangling bonds saturated by H atoms. Since the edge
atoms have 2 and the terrace atoms have 1 dangling bond each, we
require 12 H atoms to saturate all the dangling bonds on the bottom
surface. The top five Si layers
are relaxed for geometry optimization while the two lower-most Si and
the H layers are held fixed to simulate the bulk-like termination. The
wavefunctions are expanded in a plane wave basis set with a cutoff
energy $|\vec{k} + \vec{G}|^2 \le 250$ eV. This leads to an absolute
convergence of the energy as we found in case of the ideal 
surface (defined later).
The Brillouin zone (BZ)
integration is performed within a Monkhorst-Pack (MP) scheme using 4
inequivalent k-points. 
Ionic potentials are represented by Vanderbilt-type
ultra-soft pseudopotentials~\cite{ultrasoftpp} 
and results are obtained using generalized
gradient approximation (GGA)~\cite{pw91} for the exchange-correlation potential.
Preconditioned conjugate gradient is used for wavefunction
optimization and a conjugate gradient for ionic relaxations. The z
axis is taken perpendicular to the Si(211) surface, while x and y axes
are along [$01\bar{1}$] and [$1\bar{1} \bar{1}$] respectively. All our
calculations are performed using the VASP code~\cite{vasp}.

\section{Results}
\label{sec:results}
\subsection{Clean surface}
\label{subsec:clean}

We performed a total energy calculation using the above parameters for
an ideal surface where all the Si atoms are fixed at their
bulk-terminated positions (cubic lattice constant 5.43 \AA).
We have studied the clean surface here even though part of the
calculations were already done in Ref.~\cite{grein}, the reason being
that we need information about the clean system before we can do
adsorption studies. Moreover, a comparison makes it clear that the
present calculations have a larger energy cutoff (${\rm E_{cut}}$). 
The follwoing two
tables show that the cutoff energy and k-mesh used here lead to
absolute convergence of energy. Table~\ref{table:ecut} shows the total
binding energy of the clean surface, defined as the difference between the
total energy of the system and the total energy of all its isolated
constituent atoms, for different ${\rm E_{cut}}$ with a ($2\times 2\times 1$)
MP k-mesh (4 irreducible k-points). The difference between the binding
energies with ${\rm E_{cut}}=250$ eV and $300$ eV is $4 \times
10^{-4}$ eV per atom, which is beyond the limit of accuracy of the
present calculations ($\sim 1-2 $ meV per atom). 
This indicates that ${\rm E_{cut}}=250$ eV is
enough to give absolute energy convergence. With this information, in
table~\ref{table:kmesh} we show the total binding energy of the system
for different k-meshes with ${\rm E_{cut}}=250$ eV. We find that the
difference in the total binding energies with ($2\times 2\times 1$) and
($4\times 4\times 1$) k-meshes is $8\times 10^{-4}$ eV per atom, which
is again beyond the limit of accuracy for the present calculations. 
Therefore we believe that though they do not differ significatly,
the present results would be more reliable than those in ref.~\cite{grein} 
where they do.

\begin{figure}[h]
\includegraphics{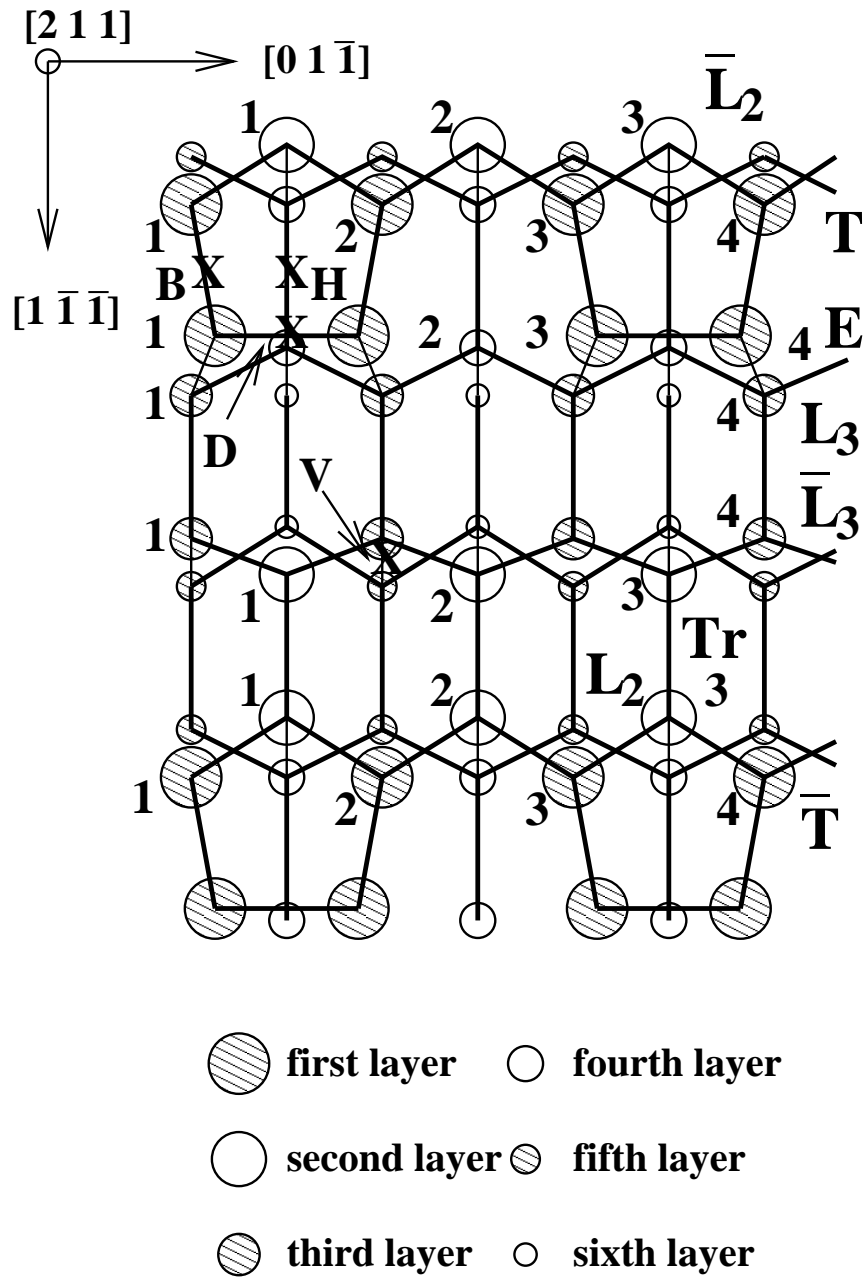}
\caption{Top view of the $(2\times 1)$ reconstructed Si(211) surface.
The two-fold coordinated edge atoms dimerize to lower the energy. The
four symmetry points on this surface at which adsorption studies have
been done are marked. Other row of atoms in the second and third layers
have been marked by ${\rm L_2}$, ${\rm \bar{L}_2}$, ${\rm L_3}$ and 
${\rm \bar{L}_3}$ which need to be identified for noting the bond lengths
and bond angles in Table~\ref{table:bla_2X1}. Note that ${\rm L_2}$ and 
${\rm \bar{L}_2}$ rows are symmetry equivalent and so are T and 
${\rm \bar{T}}$.}
\label{fig:2x1}
\end{figure}

Because of the two
dangling bonds at each edge atom, they can readily dimerize
along the [$01\bar{1}$] direction. In the optimized geometry, the
Si-Si dimer distance is found to be 2.45 \AA, a little larger than
what is found on the Si(001) surface. Because of this dimer formation, 
the surface becomes $(2\times 1)$ reconstructed. A top view of this
$(2\times 1)$ reconstructed surface is shown in Fig.~\ref{fig:2x1}. 
This doubling of unit cell along
[$01\bar{1}$] has been seen in LEED experiments, though this effect
was found to be `weak'~\cite{wang}. In addition to forming dimers, 
the edge Si atoms move up by 
$\sim 0.6$ \AA~ along the [211] direction relative to the 
corresponding terrace atoms. In STM experiments by the same
authors, the extended ordered atomic rows along
[$01\bar{1}$] were found to be made of two asymmetric thin lines close
to one another, one being positioned at a slightly greater height compared
to the other. Presumably, it is the difference in height between the
terrace and edge atoms that was seen in STM. However, the $(2\times )$
reconstructions along the [$1\bar{1} \bar{1}$] direction reported in some
experiments (the unit cell length being doubled to 19.82 \AA) 
cannot appear in our calculations as we include only two atoms along
this direction with a unit cell length of 9.41 \AA.
The dimerization of the edge Si atoms leads to an energy
gain of 2.1 eV per dimer compared to the ideal surface. Various bond lengths 
and bond angles for the $(2\times 1)$ reconstructed surface are given in
Table~\ref{table:bla_2X1}. A comparison with Ref.~\cite{grein} shows that
the Si-Si dimer distance is marginally greater in our calculation. That is why 
T(4)--E(4)--L$_3$(4) bond angle is a little smaller. It is 
noted that the $(2\times 1)$ reconstructed surface is metallic in
character.

\begin{figure}[h]
\includegraphics{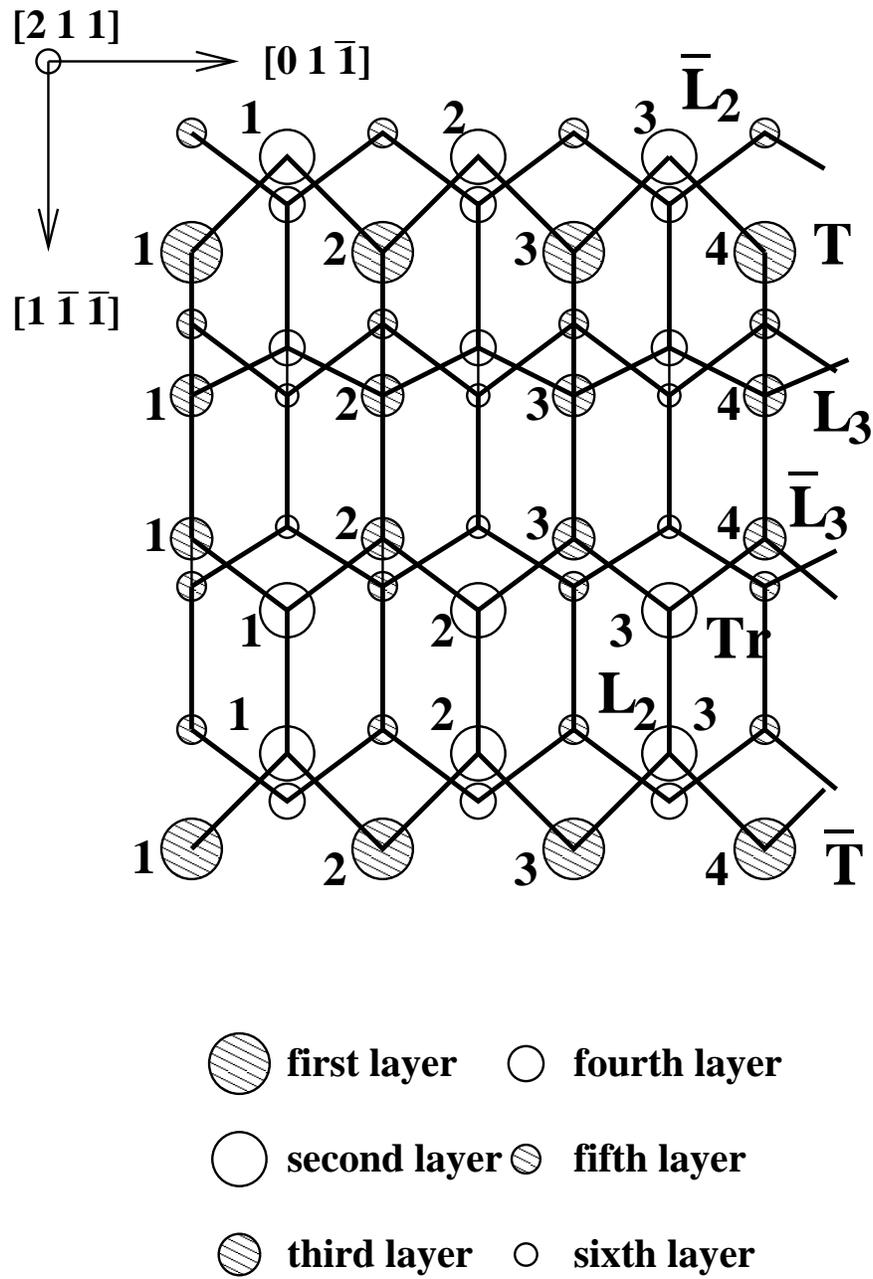}
\caption{Top view of a $(1\times 1)$ rebonded Si(211) surface. The
edge atoms are missing. The top layer atoms bind with the
corresponding third layer atoms retaining the $(1\times 1)$ symmetry
of the surface. The second and third layer rows have been identified
for the bond lengths and bond angles in Table~\ref{table:bla_1X1}.}
\label{fig:1x1}
\end{figure}

An alternate way of looking at the Si(211) surface has been suggested
in which the edge atoms are eliminated~\cite{chadi,grein}. In fact, there are
experiments that report at least parts of the surface without these
edge atoms~\cite{berghaus}. This alternate surface allows a rebonding while
maintaining the $(1\times 1)$ symmetry~\cite{chadi,pandey,grein}. 
We have also seen
this rebonded $(1\times 1)$ surface while relaxing the alternate
Si(211) surface without the edge atoms. The terrace atoms on the
surface bond with the corresponding atom along $[1\bar{1} \bar{1}]$
in the third layer. A model for
this rebonded surface is shown in Fig.~\ref{fig:1x1}. The bond lengths 
and angles for this surface are given in Table~\ref{table:bla_1X1}.
It is instructive to compare the surface energies for these two
reconstructions. Surface energy has been defined in the following way
by Chadi~\cite{chadi},
\begin{equation}
E_{surf} = E_{tot}(N) - NE_0
\end{equation}
\noindent where $E_{surf}$ is the surface energy (generally positive), $E_{tot}(N)$ is the
total energy of a system of N atoms with an exposed surface, and $E_0$
is the total energy per Si atom in the bulk. 
This is easy to see, as any difference
between two systems--one with an exposed surface and
the other an infinite bulk--will arise from the surface effects and is
a measure of the energy required to create the surface. Grein used this definition to
calculate the surface energies of the $(2 \times 1)$ and rebonded
$(1 \times 1)$ surfaces in his calculation~\cite{grein}. Note that 
in that calculation, the dangling bonds in the lower-most Si layer were
not saturated by H. However, it was possible to take care of the
effects of these dangling bonds while calculating
the surface energy of the reconstructed surface at the top.

In the present calculations, in contrast, we have saturated the
dangling bonds at the bottom by H. This allows one to simulate a
bulk-like termination at the bottom with a relatively small slab
thickness. Unfortunately, this also requires us to know the
contribution of the Si-H bonds to the total energy of the system if we
wish to calculate the surface energies from these calculations.
Although we are unable to obtain the surface energies of the $(2\times 1)$
and rebonded $(1 \times 1)$ surfaces separately, our main interest is to calculate the
difference in the surface energies of these two reconstructions. 
This indeed can be done, since the
unknown contribution from the Si-H bonds is exactly the same in the
two systems and cancels out exactly. Another point to remember
while calculating the surface energy difference between the two
reconstructions is that the two systems have different number of
atoms--the system with $(2\times 1)$ reconstructed surface has 4
surface atoms more than the rebonded $(1\times 1)$ surface. Assuming that
the $(2\times 1)$ and rebonded $(1\times 1)$ systems have N and (N$-$4) 
Si atoms forming the slabs respectively, from the above argument we can
write,
\begin{equation}
E_{surf}(2\times 1) = E_{total}(2\times 1,N) - NE_0 + E_{Si-H}
\label{eqn:surf2x1}
\end{equation}
\noindent The surface energy of a rebonded $(1\times 1)$ surface can
be written as 
\begin{equation}
E_{surf}(1\times 1) = E_{total}(1\times 1,N-4)-(N-4)E_0 + E_{Si-H}
\label{eqn:surf1x1}
\end{equation}
\noindent where $E_{Si-H}$ is the unknown contribution to the energy 
from the Si-H bonds.  
Taking the difference of
eqns.(~\ref{eqn:surf2x1}) and (~\ref{eqn:surf1x1}), 
and using the values of the energies from our calculations we find,
\begin{equation}
E_{surf}(2\times 1) - E_{surf}(1\times 1) = -1.9356 \;\; {\rm eV}
\end{equation}
The surface area of the Si slab in our supercell is 144.914 \AA$^2$.
This gives an energy difference of 
0.013 eV/\AA$^2$ between the two surfaces. It should be noted here that
like in Ref.~\cite{grein}, the energy difference between the two surface
reconstructions is rather small and is at the level of accuracy of these
calculations. Hence, depending on the actual
experimental conditions, one may end up with either surface.

\begin{figure}[h]
\includegraphics{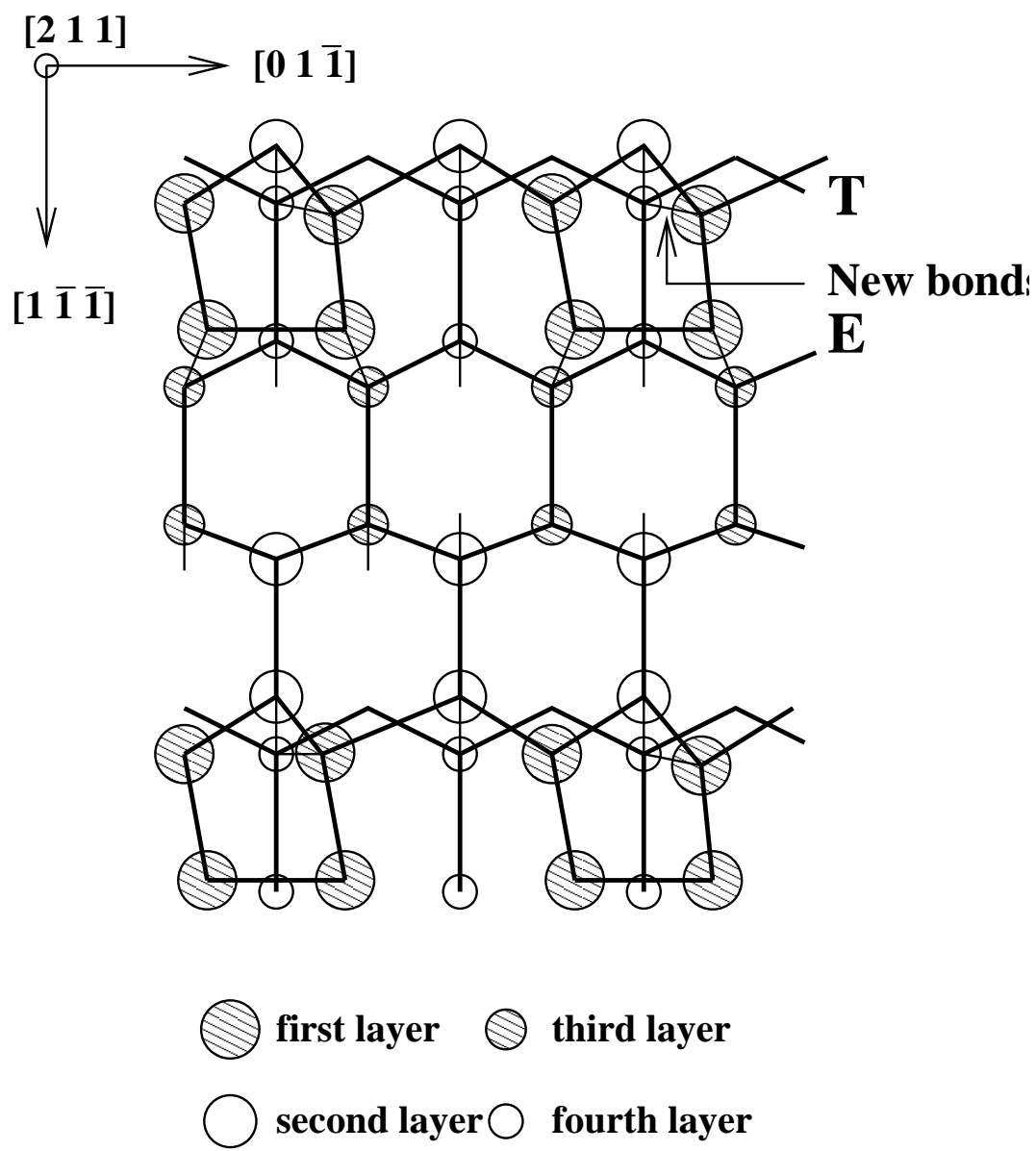}
\caption{The proposed $\pi$-bonded chain reconstruction of Si(211)
surface. The top layer {\bf T} atoms bind with the corresponding
fourth layer atoms. Only the top four layers have been shown for
clarity.}
\label{fig:pi-chain}
\end{figure}

Another reconstruction that was proposed as a possibility for the
Si(211) surface is a $\pi$-bonded chain formation as on the Si(111)
surface. A model for $\pi$-bonded chain reconstruction on the Si(211)
surface is shown in Fig.~\ref{fig:pi-chain}, 
in which the top layer terrace atoms 
dimerize with the corresponding fourth layer
atoms~\cite{kaxiras}. We investigate this possibility extensively. First, we
push down the top layer terrace atoms by $\sim 1.1$ \AA~ so that they
are at the same height as the second layer atoms to facilitate any
possible dimerization, and relax the structure. In the optimized
structure, the top layer terrace atoms are pushed back up to their
original height and the edge atoms form dimers along the [$01\bar{1}$]
direction, leading to the $(2\times 1)$ structure.
Next, we start from an initial
geometry in which the top layer terrace atoms are pushed
down substantially so that the distance between them and the corresponding
fourth layer atoms is $\sim 2.3$ \AA, close to the Si-Si bond length. Then, as a
first step, these atoms forming the `dimer' are fixed along with 
the two lower-most Si layers and
the H layer, while all the other atoms are relaxed. The energy of
this configuration is found to be higher than that in a $(2\times 1)$
reconstructed surface. Starting with this intermediate
geometry, all the atoms in the top 5 layers are allowed to move
(including the `dimers' that were held fixed) for further structural
optimization. The top layer atoms again move up.
The surface ends up in the same $(2\times 1)$ reconstruction with 
the same energy as obtained starting from an ideal surface. Thus we conclude
that the $\pi$-bonded chain is not a stable reconstruction for
this surface.

\subsection{Arsenic adsorption}
\label{subsec:As}

In this section we discuss our results of As adsorption on the Si(211)
surface. Though it has been reported that the clean Si(211) 
can form a rebonded
$(1\times 1)$ surface where the edge atoms are missing, none of the adsorption
experiments on this surface report this. Also, as noted above, our calculations
give a slightly lower surface energy for the $(2\times 1)$ surface. Hence we
have restricted our adsorption studies only to the ideal and the
$(2\times 1)$ reconstructed surfaces.

\subsubsection{As on ideal Si(211)}
A stepped surface like Si(211) has many
inequivalent atoms on the surface and hence offers a number of
possible adsorption sites.
We study the adsorption of an As atom on six different
symmetry sites on the ideal Si(211) surface. These symmetry points
are labeled in Fig.~\ref{fig:ideal}. The {\bf T}, {\bf E} and {\bf
Tr} sites are on top of the terrace, edge and trench Si atoms marked
in the figure. The trench atoms are the second layer Si atoms that are
also 3-fold bonded. The bridge site ({\bf B}) is on top of the bond between the
surface terrace and edge atoms, while the hollow site ({\bf H}) is in the middle
of the rectangle on the surface formed by two terrace and
the two corresponding edge atoms. The valley site ({\bf V}) is halfway between two
second layer trench atoms. We have selected a set of plausible
adsorption sites from our intuitive notions
of chemical bonding.  

\begin{figure}[h]
\includegraphics{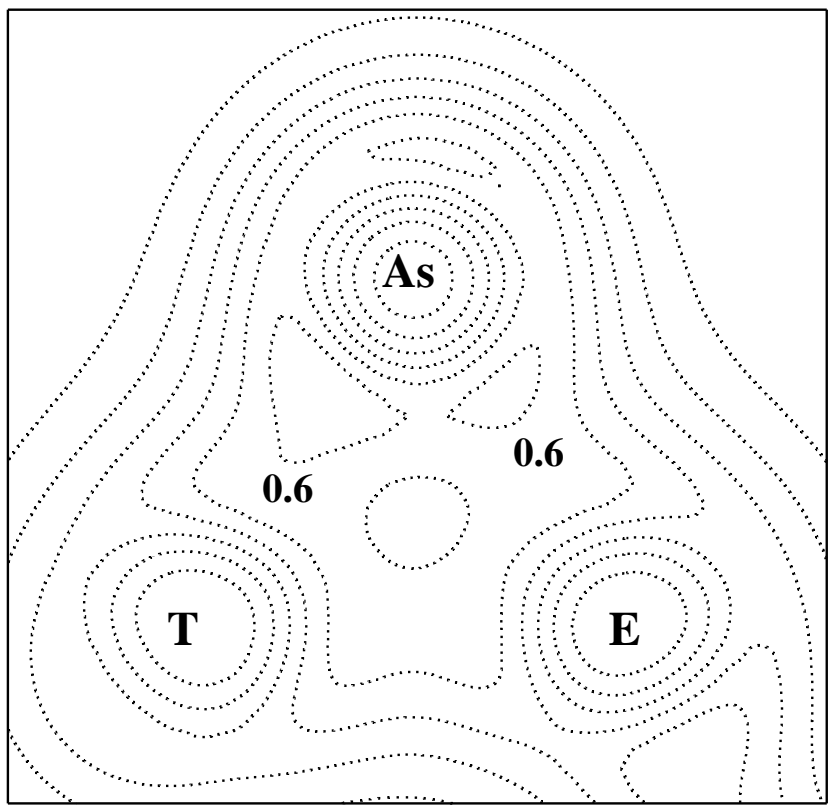}
\caption{Charge density contour plot in the Si-As-Si plane (y-z plane in 
our calculation) for an isolated As adatom at the {\bf B} site of an ideal
Si(211) surface. The asymmetry in bonding of the As with the terrace and
edge Si atoms is apparent.}
\label{fig:idlAs_rhor}
\end{figure}

The binding energies of an As atom at
various sites and its distance from the nearest Si atoms(s) are given in
Table~\ref{table:As}. The {\bf B} site turns out to be energetically
the most favorable one for As adsorption. This is followed by the
{\bf H}, {\bf V}, {\bf E}, {\bf Tr} and {\bf T} sites.
This sequence of binding energies can be rationalized in the
following way. At the {\bf B} site, the As can bind to the terrace
and edge Si. With the edge Si, which has two dangling bonds left, it
can form a double bond. The asymmetry in the binding of the As with
two surface Si atoms at the {\bf T} and {\bf B} sites is clearly seen
in the charge density plot in Fig.~\ref{fig:idlAs_rhor}. 
This satisfies the dangling bonds of the
terrace and edge Si atoms. The As is 3-fold bonded and its remaining
two electrons can from a lone pair. At the {\bf H} site, the As forms
four weak bonds with the four neighboring Si atoms--two terrace and
two edge--as can be seen from the larger Si-As bond lengths. However,
formation of four bonds reduces the number of dangling bonds. Arsenic is
left with one dangling bond and the edge Si atoms with one dangling
bond each. The terrace Si's have their bonds satisfied. This
reduction of dangling bonds lead to quite large binding energy at this
site. On top of the edge, terrace and trench atoms, the As can bind
with only one Si atom. Thus the binding energies are unfavorable at
these sites. Out of these three, the binding energy at the {\bf E}
site is the highest. This is presumably because As forms a double bond
with the edge Si whereby satisfying its dangling bonds and two of its
own. At the {\bf T} and {\bf Tr} sites, however, it can form only one
bond with the corresponding Si atoms as they have only one dangling
bond each. This conjecture is supported by the fact that at the {\bf
E} site, the As-Si bond length is the shortest indicating a stronger
binding. On the other hand, at
the {\bf V} site, the As can form bonds with two neighboring trench Si
atoms, which have their remaining dangling bonds satisfied in the
process. This makes the binding energy at the {\bf V} site comparable
to that at the {\bf B} and {\bf H} sites.

\subsubsection{As on Si(211)-$(2\times 1)$}

We now discuss our studies of As adsorption on the $(2\times 1)$
reconstructed Si(211) surface. We have already seen in case of the
ideal surface that the sites on top of various surface atoms, where
the As binds to only one Si atom, are not favorable energetically. So we
do not consider those sites any more. We study the adsorption of As on
the {\bf B}, {\bf V} and {\bf H} sites. In addition, we consider
As adsorption on top of the surface Si-Si dimer ({\bf D} site, see
Fig.~\ref{fig:2x1}). For these
calculations, the planar position of the As adatom is held fixed
while it is allowed to relax its height. The top five Si layers are
fully relaxed as before. 

\begin{figure}[h]
\scalebox{0.6}{
\includegraphics{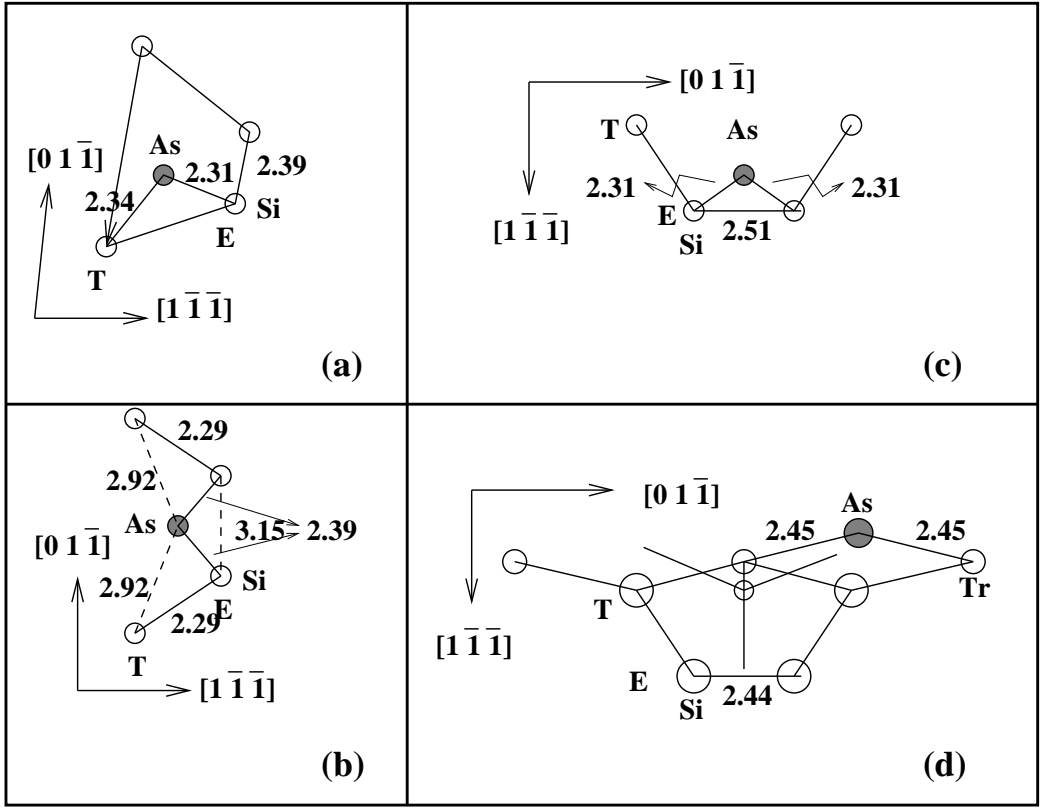}}
\caption{The local geometry around the adsorbed As atom at the {\bf B}
(a), {\bf H} (b), {\bf D} (c) and {\bf V} (d) sites of the $(2\times
1)$ reconstructed Si(211) surface. Please note that the direction of
view is slightly different in different cases. All the distances are
given in \AA.}
\label{fig:As-geom}
\end{figure}

On the $(2\times 1)$ reconstructed surface, the {\bf V} site turns out
to be the most favorable one for an As adatom with a binding energy 
$E_b^V=4.84$ eV. This is followed by the {\bf B}, 
{\bf D} and {\bf H} sites with $E_b^B=4.69$ eV, $E_b^D=4.32$ eV
and $E_b^H=3.81$ eV. The geometries of the surface around the As
adatom in these four cases are shown in Fig.~\ref{fig:As-geom} . 
At the {\bf V} site, As binds with the two neighboring second layer Si atoms.
Similarity of As adsorption at this site to that on the Si(001)
surface is noticeable. A monolayer of As adsorbed on
Si(001) has As-Si bond length equal to 2.44 \AA~\cite{uhrberg}. 
As adsorbed on the
{\bf V} site has the same As-Si bond length. 
At the {\bf B} site, the As binds to two 
surface Si atoms as on an ideal surface.
However, it cannot from a double bond with the edge Si as on 
an ideal surface, since the edge Si is already
three fold coordinated because of the Si-Si dimerization. Binding with
the As satisfies the dangling bonds of the terrace and edge Si atoms.
In spite of having the same dangling bond density, As adsorbed at the {\bf
V} site has a higher binding energy probably because of the optimal length
of the As-Si bond, which is much shorter ($\sim 2.3$ \AA) in case of
As on the {\bf B} site.

Interestingly, presence of the As adatom at the {\bf B} site reduces
the Si-Si dimer distance marginally to 2.39 \AA. At the {\bf D} site,
the As binds with the two edge Si atoms forming the dimer. However,
presence of the As atom in this case increases the Si-Si dimer
distance marginally to 2.52 \AA, costing energy. This makes the binding
energy on this site slightly lower than that at the {\bf B} site. At the
{\bf H} site, As prefers to form bonds of optimal length with the
two edge Si atoms and does not bind with the terrace atoms as can be
seen from the interatomic distances shown in Fig.~\ref{fig:As-geom}. 
Apparently this
cannot be achieved by keeping all the Si-Si bonds intact.
Consequently, the Si-Si dimer bond is broken making the Si-Si distance
3.15 \AA~ and costing the Si-Si binding energy. This is in contrast to
the behavior of $(2\times 1)$ reconstructed Si(001) surface on which As
and Te are known to break the Si-Si surface dimers adsorbing on
top of the dimers. This breaking of the Si-Si dimers makes the binding
at this site the least beneficial energetically. However, it must be
noted that though the binding energy at the hollow site is quite low
on the $(2\times 1)$ surface compared to an ideal one, the total
energy is still lower in the former case. 

We have also studied the effects of a higher coverage of As on the
$(2\times 1)$ reconstructed Si(211) surface. Although the {\bf V} site
turned out to be the most favorable one for an isolated As adsorption,
we put the As atoms at the next most favorable site, namely the {\bf
B} site, at higher coverages. This is because, at the {\bf V} site,
the As adatom does not have any major effect on the geometry of the
surface apart from binding with the two neighboring {\bf Tr} atoms.
So, {\it a priori} it seems that at higher coverage, As atoms, if
adsorb at the alternate {\bf V} sites, would bind with the {\bf Tr} atoms
without causing any change in the reconstruction of the surface. 
Therefore, at a coverage of 0.5 ML that we study, we put four As atoms
on the four {\bf B} sites of the $(4\times 2)$ surface supercell that
we use.

\begin{figure}[h]
\scalebox{0.6}{
\includegraphics{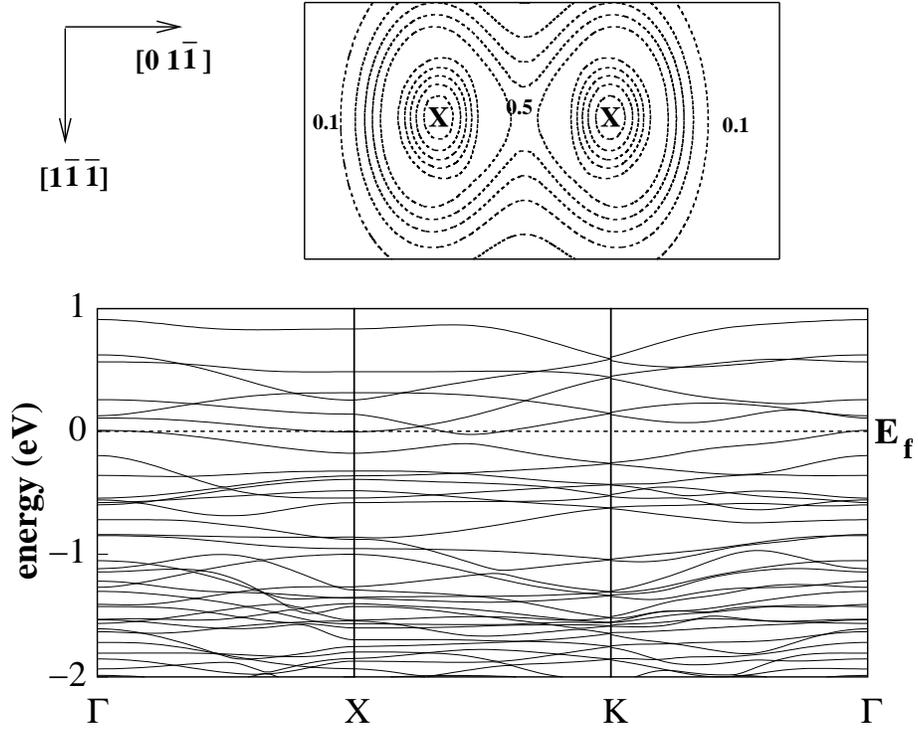}}
\caption{The top panel shows the charge density plot in the plane 
of the As atoms for 0.5 ML coverage. The maximum charge density 
halfway between the two As
atoms at the {\bf B} sites indicate dimerization. The bottom panel
shows the band structure of the same system.}
\label{fig:As-rhor}
\end{figure}

At 0.5 ML coverage, we find a binding energy of 5.28 eV per As adatom.
This is larger than the binding energy of a single As adatom on a {\bf
B} site. With As adatoms on all the {\bf B} sites, the Si-Si dimer
distances decrease to 2.31 \AA. This effect was already seen in
presence of a single As atoms at a {\bf B} site. Presence of more As
adatoms enhances this effect further. Energy is gained because of
stronger binding of the surface Si atoms. Further energy
is gained due to dimerization of the As adatoms. That the As adatoms
indeed dimerize at 0.5 ML can be seen from the charge density plot in
the plane of the As atoms shown in Fig.~\ref{fig:As-rhor}. This figure
also shows the band structure of the system along directions parallel to
the surface. The system is seen to be metallic in character.
In the final optimized
structure, the Si-As and As-As distances are 2.44 \AA~ and 2.55 \AA~
respectively, exactly equal to the corresponding numbers for As
adsorption on the Si(001) surface. 
The surface still retains the $(2\times 1)$ reconstruction at this coverage,
but the origin is due to the dimerization of the As adatoms.
\subsection{Te adsorption on Si(211)}
\label{subsec:Te}

In this subsection, we discuss our studies of Te adsorption on the
Si(211) surface. As in the case of As adsorption, we study adsorption
of Te on the ideal and the (2$\times$1) reconstructed surfaces.

\subsubsection{Te on ideal Si(211)}

As in the case of As adsorption, we have studied Te adsorption at six
special symmetry points on the ideal Si(211) surface. These sites can 
be grouped into two categories in terms of the
nature of binding and the binding energies. 
Three sites, namely the {\bf H}, {\bf V} and
{\bf B} sites have high binding energies which are close to each other.
The {\bf T}, {\bf E} and {\bf Tr} sites, on the other had, have
relatively low binding energies. The reason is similar to the case of
As. At the {\bf H}, {\bf V} and {\bf B} sites, the Te binds to four,
two and two Si atoms respectively. This reduces the number of dangling
bonds leading to high binding energies. Why the binding energies at
the {\bf B} and {\bf V} sites are lower than that at the {\bf H} site
is not obvious. It appears that the 
bond-bending energy of the edge Si atom for a Te at {\bf B} is rather high. For
an As at {\bf B}, further energy can be gained by forming a double
bond with the edge atom which is not possible for a Te adatom. It
should also be mentioned that the binding energies at the {\bf V} and
{\bf B} sites are very close and their difference is 
within the error margins of our
calculation. This is in agreement with the experimental observation of
Dhar {\it et al.}~\cite{dhar} mentioned before. However, they have no precise
information of the binding sites because of the nature of their
experiments.
For a Te on top of the {\bf T}, {\bf E} and {\bf Tr} Si
atoms, the Te adatom can bind to only one Si atom and thereby have low
binding energies. The binding energies and the distances from the
nearest Si atoms are given in Table~\ref{table:Te}.

\subsubsection{Te on Si(211)-$(2\times 1)$}

\begin{figure}[h]
\scalebox{0.6}{
\includegraphics{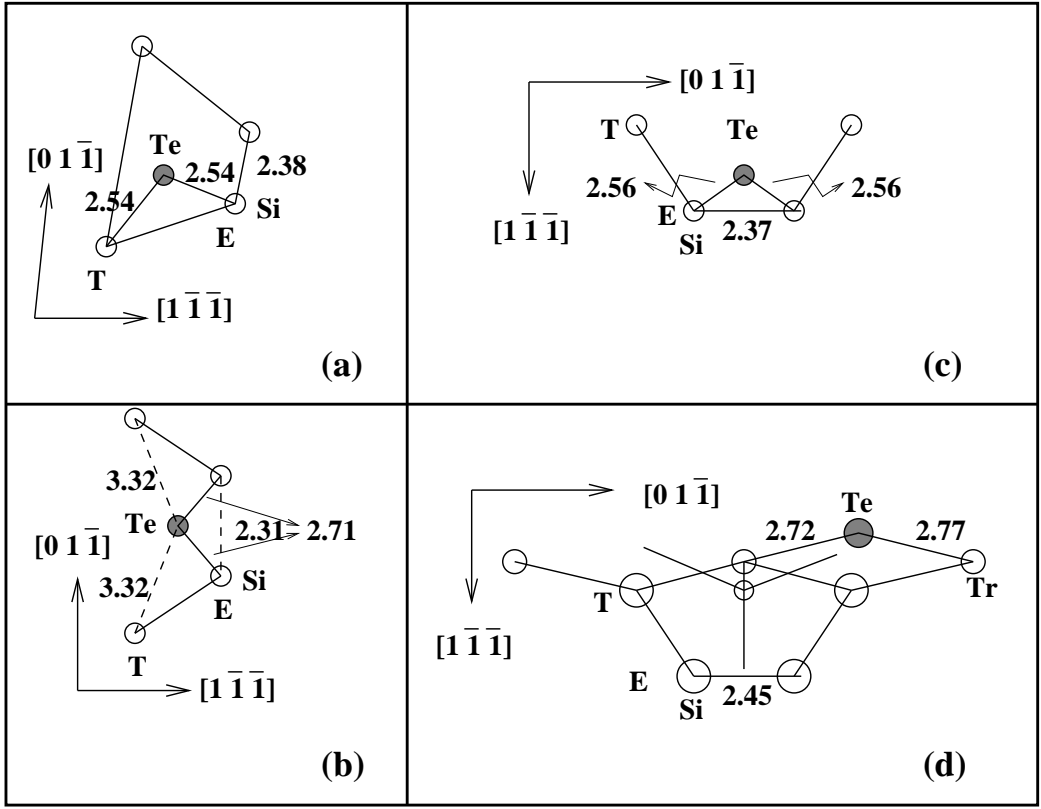}}
\caption{Local geometry around the Te adatom at various sites on a
$(2\times 1)$ reconstructed Si(211) surface: (a) {\bf B}, (b) {\bf H},
(c) {\bf D} and (d) {\bf V} sites.}
\label{fig:Te-geom}
\end{figure}

Now we discuss our studies of Te adsorption on the $(2\times 1)$
reconstructed Si(211) surface. Again, we study adsorption only on the
four sites where the Te adatom can bind to more than one Si atoms,
i.e., the {\bf B}, {\bf H}, {\bf V} and the {\bf D} sites. The {\bf B}
site is found to be energetically the most favorable one for Te
adsorption with a binding energy of $E_b^B=4.09$ eV, which is closely
followed by the {\bf V} site with $E_b^V=4.02$ eV. The {\bf D} and
{\bf H} sites have relatively lower binding energies with
$E_b^D=3.32$ eV and $E_b^H=2.62$ eV respectively. It should be
mentioned here that the maximum binding energy values fot Te on
Si(211) found in our calculation is slightly different from that
obtained by Dhar {\it et al.}.~\cite{dhar} However, that is not
surprising given the fact that ours is a complete quantum mechanical
calculation while their estimate is based on purely classical
considerations. Unlike in case of
As, the Si-Si dimer is not broken for a Te adsorption on the {\bf H}
site. If the As-Si bond is stronger than the Si-Si bond, the surface
Si dimers can break in favor of the Si atoms forming double bonds
with the adsorbed As thereby gaining energy. This is not possible in
case of Te which has a valence of 2 and hence the Si dimer remains
intact. The geometries around the adsorbed Te atom in these four cases
are shown in Fig.~\ref{fig:Te-geom}.

\begin{figure}[h]
\scalebox{0.6}{
\includegraphics{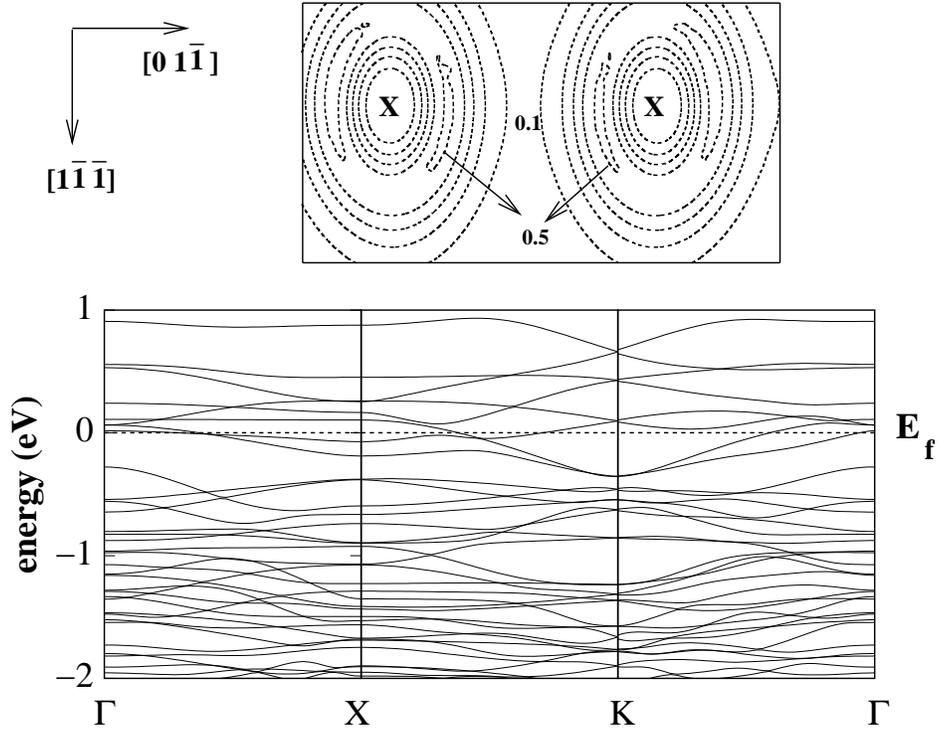}}
\caption{The top panel shows the charge density contour plot in the 
plane of Te atoms for 0.5 ML coverage. Clearly, there is no bonding between 
the two Te atoms at the {\bf B} sites. The bottom panel is the band structure
of the same system.}
\label{fig:Te-rhor}
\end{figure}

We have also studied 0.5 ML Te adsorption on the $(2\times 1)$
reconstructed Si(211) surface. Since {\bf B} site is the most
favorable for Te adsorption, it is reasonable to put the additional Te
atoms on the {\bf B} sites as well. Hence, we have studied the effects
of four Te adatoms at the four {\bf B} sites in our $(4\times 2)$
surface supercell just as in the case of As. In marked contrast to As
adsorption, the Te atoms do not dimerize as can be seen from the
charge density contour plots in the plane of Te atoms shown in
Fig.~\ref{fig:Te-rhor}. This figure also shows the band structure of
the 0.5 ML Te covered surface. As can be seen, this system is also
metallic in character.
The binding energy of 4.09 eV per Te adatom is almost the same as for
a single Te adatom (4.08 eV). This is consistent with the fact that
there is no additional energy gain, unlike the case of As, from adatom
dimerization. Moreover, it is also consistent with the chemical
notion that Te, having the outermost electronic configuration of
$5s^25p^4$, is divalent and can not dimerize after binding with two
neighboring Si atoms. This absence of Te-Te dimerization is exactly
similar to the case of 1 ML Te adsorption on the Si(001)
surface~\cite{sen}. However, in case of the Si(211) surface, the
$(2\times 1)$ reconstruction of the substrate is retained because the
edge Si atoms retain their dimerization. In the optimized structure,
the Si-Si dimer distance on the surface
is found to be 2.35 \AA; and the Si-Te distance is 2.56
\AA. The Si-Te distance is very close to the sum of the atomic radii 
of Si ($\sim 1.17$ \AA) and Te ($\sim 1.32$ \AA) and the
value for the same quantity found for Te adsorption on the Si(001) surface.
Note that we have not tried to study coverages beyond 0.5 ML because of
the additional complication of sub-surface adsorption for coverages
greater than 0.6 ML, as seen in ref~\cite{dhar}.

\section{Conclusion}
\label{sec:conclusion}

We have performed an extensive plane-wave, pseudopotential
density functional calculations for the electronic and structural
properties of the clean, and As and Te covered Si(211) surface. The
clean surface readily forms a $(2\times 1)$ reconstruction through
dimerization of the edge atoms along the [$01\bar{1}$] direction. A
rebonded $(1\times 1)$ surface is also found, 
but the $\pi$-bonded chain is not a
stable reconstruction for this surface. The two-fold bridge site is found
to be the most favorable one for an isolated As adatom, whereas the
four-fold hollow site is the most favorable one for a Te on an ideal
Si(211) surface. On the
$(2\times 1)$ reconstructed surface, the valley site is the most
favorable one for As while the bridge site is the most favorable one
for an isolated Te. Just as suggested by the experiments on Br and Te
adsorption on the Si(211) surface, there are more than one points on
the surface with very close binding energy values for both As and Te.
At 0.5 ML coverage, when there is a line of adatoms along
[$01\bar{1}$] at the bridge sites, the As adatoms dimerize on the
Si(211) surface while Te adatoms do not. In this respect, their
behavior is very similar to that on the Si(001) surface. Both 0.5 ML
As and Te covered surfaces are found to be metallic in character.

\newpage

\clearpage

\begin{table}
\caption{Total binding energy (${\rm E_b}$) of the clean Si(211) surface for
different ${\rm E_{cut}}$ with 4 irreducible k-points.}
\begin{tabular}{cr}
\hline \hline
${\rm E_{cut}}$ (eV)   &    ${\rm E_b}$ (eV)  \\   \hline
100           &  $-316.598285$  \\

150           &  $-323.238185$  \\

200           &  $-324.794424$  \\

250           &  $-325.044874$  \\

300           &  $-325.075853$  \\    \hline
\end{tabular}
\label{table:ecut}
\end{table}

\begin{table}
\caption{Total binding energy (${\rm E_b}$) of the clean Si(211) surface for
different MP k-meshes with ${\rm E_{cut}}=250$ eV.}
\begin{tabular}{ccr}
\hline  \hline
k-mesh              & irreducible k-points &     ${\rm E_b}$ (eV)   \\  \hline

$1\times 1\times 1$ &   1                  &     $-323.661536$  \\

$2\times 2\times 1$ &   4                  &     $-325.044874$   \\

$4\times 4\times 1$ &   10                 &     $-324.990027$    \\  \hline
\end{tabular}
\label{table:kmesh}
\end{table}

\begin{table}
\caption{Selected bond lengths and bond angles for a $(2\times 1)$ surface}
\begin{tabular}{lc}

\hline \hline 
Atoms                   &      Bong length (\AA) or angle ($^0$) \\  \hline
                                                         
T(4)--E(4)              &      2.26    \\
                                       
E(3)--E(4)              &      2.45    \\
                                       
E(4)--L$_3$(4)          &      2.37    \\
                                       
L$_3$(4)--${\rm \bar{L}_3(4)}$            &      2.35    \\
                                       
${\rm \bar{L}_3(4)}$--Tr(3)           &      2.32    \\
                                       
Tr(3)--L$_2$(3)            &      2.36    \\
                                       
 ${\rm \bar{L}_2(3)}$--T(4)--E(4)   &     88.67    \\
                                       
 T(4)--E(4)--L$_3$(4)     &    105.04    \\
                                       
 E(4)--L$_3$(4)--${\rm \bar{L}_3(4)}$     &    108.42    \\   \hline

\end{tabular}
\label{table:bla_2X1}
\end{table}

\begin{table}
\caption{Selected bond lengths and bond angles for the $(1\times 1)$ surface}
\begin{tabular}{lc}
\hline    \hline   
Atoms                    &   Bond length (\AA) and angle ($^0$) \\  \hline
                                                    
 T(4)--L$_3$(4)                  &     2.46            \\
                                                    
 L$_3$(4)--${\rm \bar{L}_3(4)}$                  &     2.52            \\
                                                    
 ${\rm \bar{L}_3(4)}$--Tr(3)                 &     2.34            \\
                                                    
 Tr(3)--L$_2$(3)                  &     2.41            \\
                                                    
 L$_2$(3)--${\rm \bar{T}(4)}$             &     2.43            \\
                                                    
 ${\rm \bar{L}_2(3)}$--T(4)--L$_3$(4)        &   101.80            \\
                                                    
T(4)--L$_3$(4)--${\rm \bar{L}_3(4)}$           &   137.56            \\
                                                    
L$_3$(4)--${\rm \bar{L}_3(4)}$--Tr(3)           &   120.45            \\   \hline

\end{tabular}
\label{table:bla_1X1}
\end{table}

\begin{table}
\caption{Distances of the As adatom at various symmetry points on the
surface from the neighboring Si atoms of an ideal surface. The corresponding
binding energies are also given.}
\label{table:As}
\begin{tabular}{lccccccc}
\hline   \hline
As & \multicolumn{6}{c}{distance (\AA) from which Si} \\ \cline{2-7}
position & E(1) & E(2) & T(1) & T(2) & Tr(1) & Tr(2) & $E_b$ (eV) \\  \hline
{\bf B}  & 2.30 & ---  & 2.30 & ---  & ---  & ---  & 5.37 \\ 
{\bf H}  & 2.61 & 2.61 & 2.61 & 2.61 & ---  & ---  & 5.30 \\
{\bf V}  & ---  & ---  & ---  & ---  & 2.59 & 2.59 & 5.05 \\
{\bf E}  & 2.16 & ---  & ---  & ---  & ---  & ---  & 4.55 \\
{\bf T}  & ---  & ---  & 2.24 & ---  & ---  & ---  & 3.85 \\
{\bf Tr} & ---  & ---  & ---  & ---  & 2.26 & ---  & 4.28 \\  \hline 
\end{tabular} 
\end{table}

\bigskip

\begin{table}
\caption{Distances of the Te adatom at various symmetry points on the
surface from the neighboring Si atoms of an ideal surface along with the
corresponding binding energies.}
\label{table:Te}
\begin{tabular}{lccccccc}
\hline   \hline
Te & \multicolumn{6}{c}{distance (\AA) from which Si} \\ \cline{2-7}
position & E(1) & E(2) & T(1) & T(2) & Tr(1) & Tr(2) & $E_b$ (eV) \\  \hline
{\bf B}  & 2.52 & ---  & 2.52 & ---  & ---  & ---  & 4.15 \\
{\bf H}  & 2.80 & 2.80 & 2.80 & 2.80 & ---  & ---  & 4.26 \\
{\bf V}  & ---  & ---  & ---  & ---  & 2.79 & 2.79 & 4.19 \\
{\bf E}  & 2.38 & ---  & ---  & ---  & ---  & ---  & 3.65 \\
{\bf T}  & ---  & ---  & 2.47 & ---  & ---  & ---  & 3.58 \\
{\bf Tr} & ---  & ---  & ---  & ---  & 2.46 & ---  & 3.85 \\  \hline
\end{tabular}
\end{table}

\end{document}